\documentclass[11pt]{article}
\usepackage{graphicx}
\usepackage[margin=1.25in]{geometry}
\usepackage[usenames,dvipsnames]{color}
\usepackage{url}
\usepackage[colorlinks = true,
            linkcolor = blue,
            urlcolor  = blue,
            citecolor = blue,
            anchorcolor = blue]{hyperref}

\usepackage{todonotes}

\makeatletter
\renewcommand\paragraph{\@startsection{paragraph}{4}{\z@}%
            {-2.5ex\@plus -1ex \@minus -.25ex}%
            {1.25ex \@plus .25ex}%
            {\normalfont\normalsize\bfseries}}
\makeatother
\def\Title#1{\begin{center} {\LARGE #1 } \end{center}}
\def\Author#1{\begin{center}{ \sc #1} \end{center}}
\def\Address#1{\begin{center}{ \it #1} \end{center}}

\newenvironment{Abstract}{\begin{quotation} \begin{center}
                       ABSTRACT
     \end{center}\bigskip  }{\end{quotation}}

\newcommand\snowmass{\begin{center}\rule[-0.2in]{\hsize}{0.01in}\\\rule{\hsize}{0.01in}\\
\vskip 0.1in Submitted to the  Proceedings of the US Community Study\\ 
on the Future of Particle Physics (Snowmass 2021)\\ 
\rule{\hsize}{0.01in}\\\rule[+0.2in]{\hsize}{0.01in} \end{center}}

\begin{document}

%\pubblock

\Title{Summary Report of the Topical Group on Physics Education\\
Community Engagement Frontier (CEF4/CommF4)\\
Snowmass 2021\\
}

\bigskip 

\Author{Sijbrand J. de Jong$^1$, Sudhir Malik$^2$, Randal Ruchti$^3$}
\Address{$^1$Faculty of Science, Radboud University, 6525 AJ Nijmegen, The Netherlands}
\Address{$^2$Physics Department, University of Puerto Rico Mayaguez, PR 00682, USA}
\Address{$^3$Department of Physics and Astronomy, University of Notre Dame, Notre Dame, IN 46556 USA}
\medskip

\begin{Abstract}
An essential companion to the development and advancement of the field of Particle Physics is a strong program in physics education at all levels, that can attract entry level students across the full demographic spectrum and provide them with the education, training and skills needed to advance to successful careers in Science, Technology, Engineering and Mathematics (STEM) and other fields.   This report summarizes the work of several investigative teams that have reviewed and  assessed current opportunities in physics education across K-12, undergraduate, graduate and postdoctoral domains, including national and international linkages.  From these assessments, recommendations have been put forward aimed to innovate educationally in strategic ways to strengthen ties between the research community and teachers, between the academic community and the private sector, and through both domestic and international connections.
\end{Abstract}
\snowmass
%\end{titlepage}
 
%\tableofcontents
\clearpage

\section{Introduction}   
\label{section:Introduction}

To have in place the needed educated and trained workforce for currently planned and future experiments in high energy particle physics and ancillary fields, the attraction to and education of students for a broad range of careers in STEM is both necessary and essential.  

\textbf{The Challenge:} Traditional STEM educational efforts have provided a high-quality workforce, but with rather selective demographics. And if left unchanged, the character of the workforce will likely remain stagnant and generally discouraging to the participation of women and those from underrepresented groups.  

\textbf{The Opportunity:} The Snowmass 2021 Process, with its 10-year planning and 20-year vision, offers a creative opportunity to assess the challenges, build on what is currently working very well, and frame a structure to broader opportunity for young researchers to join the exciting particle physics field through an expanded range of educational opportunities. 

\begin{figure}[htb]
\begin{center}
\includegraphics[width=1.0\hsize]{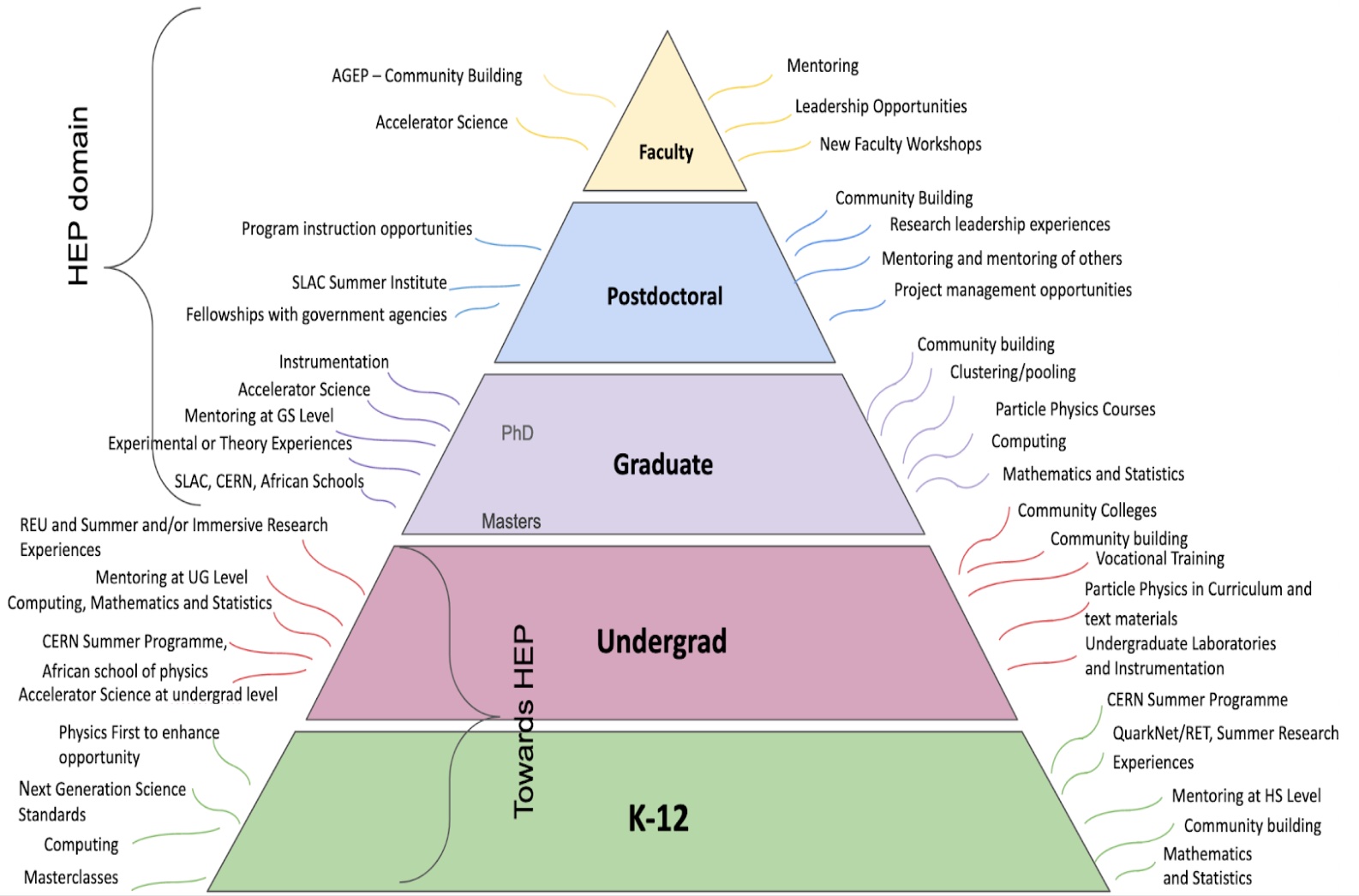}
\hfill
\end{center}
\caption{A Schematic Representation of Particle Physics Education}
\label{fig:Pyramid2}
\end{figure}

\textbf{The Approach and Organization of this Topical Group:} To identify what is working and the shortfalls, and to recommend actions to be taken, the Community Engagement Frontier Topical Group on Physics Education has viewed the physics education process in a systemic way as indicated  schematically in Figure~\ref{fig:Pyramid2}.  The schema highlights the educational process from the level of incoming K-12 students and then upward through undergraduate and graduate education, postdoctoral and faculty education.   
%%%%%%%%%%%%%%%%%%%%%%%%%%%%%%%%%%%%%%%%%%%%%%%%%%%%%%%%%%%%%%%%%%%%%%%%%
%%%%%%%%%%%%%%%%%%%%%%%%%%%%%%%%%%%%%%%%%%%%%%%%%%%%%%%%%%%%%%%%%%%%%%%%%%%
A very large and demographically diverse population of students who might consider and potentially enter STEM fields enters at the base of the pyramid.  For a variety of reasons, educational,  cultural and societal, the fraction of students that choose STEM careers and are able proceed upward to research careers becomes reduced at each step.  To assess these issues, the Physics Education Topical Group formed four Working groups (A through D)  organized by community interest, all of which have submitted contributed papers to the arXiv and include: {\it Group A. Opportunities for Particle Physics Engagement in K-12 Schools and Undergraduate Education}~\cite{cp2}; {\it Group B. Transforming U.S. Particle Physics Education: A Snowmass 2021 Study}~\cite{cp3}; {\it Group C. Broadening the Scope of Education, Career and Open Science in HEP}~\cite{cp14}; and {\it Group D. The Necessity of International Particle Physics Opportunities for American Education}~\cite{CP5}.  

\textbf{The Organization of this Report:} In this Topical Group Summary Report, we follow the schematic (and systemic) flow up the Pyramid of Figure~\ref{fig:Pyramid2} and highlight briefly important issues identified by each group including selected recommendations.   The education process arises naturally at the K-12 and undergraduate levels where the challenge is to attract and engage young students to interest in STEM and physics, and then to hold and enrich their interest and provide needed scaffolding to help sustain their upward path to research careers.  Greater depth and insight will be found in each of the associated and referenced Contributed Papers indicated above ~\cite{cp2,cp3,cp14,CP5}.   At the end of this report, we provide potential linkages of this Physics Education Topical Group with the other Topical Groups of the Community Engagement Frontier of Snowmass 2021, and conclude with a final comment. 

\section{Opportunities for Particle Physics Engagement in K-12 Schools and Undergraduate Education}
\label{section:Section2}

While many (particularly young) students might show an early interest and aptitude for science and mathematics at the elementary level, the structures are not necessarily in place to capture, nurture and develop such nascent interest.
In the contributed paper~\cite{cp2} it was found that joint activities of academia and K-12 educators and pupils should focus on how scientists develop knowledge and the essence of the knowledge acquired so far. It should connect to sister STEM fields, to their successful outreach programs and with an emphasis of a holistic view on STEM. It is important that K-12 activities address all students and teachers, adapting to their appropriate level of scientific literacy, and be mindful of biases in program designs that might filter student participation by race, gender, or socioeconomic background.
The paper~\cite{cp2} makes a number of recommendations that can be condensed into:
\begin{enumerate}
\setcounter{enumi}{0}
    \item Formation at the local level of collaborative communities ("fora") of STEM experts of all backgrounds (physicists, engineers, technicians, students and K-12 teachers and students - to form direct linkages among researchers and educators for dialog, collaboration, mutual enrichment and mentoring among members.  These fora should be responsive to local community need.
    \item These fora should not be isolated or self-isolating.  To aid in interaction and information transfer, they should be supported by a nationally, or even internationally, organised online repository for sharing resources;
    \item The fora can only be created and sustained if a minimal amount of support for coordination and logistics is made available. For the sustainability, it is important to have a steady source for this support, which can come from colleges, universities or institutes, but might also come from or be supplemented by outreach support in research grants. 
    \item\textbf{An important ingredient for sustainable fora is that the efforts of the participants are appropriately and regularly recognised.}
\end{enumerate}

\section{Educational Opportunities at the Undergraduate, Graduate and Postdoctoral Level}   
\label{section:Section3}
This critical time domain for the development and pursuit of careers in particle physics and ancillary fields has traditionally been the purview of university and laboratory groups in the US and abroad.  In the words of the engaged group of young researchers working on this topic: "Graduate school (and undergraduate school to a lesser extent) is where researchers acquire most of the technical skills required for research, develop scientific problem solving abilities, learn how to establish themselves in their field, and begin developing their career. It is unfortunate, then, that the skills gained by physicists during their formal education are often mismatched with the skills actually required for a successful career in physics."   These words suggest that there is clearly work to do to strengthen this key educational domain.
\subsection{Online Survey of Educational Experiences in the HEP Community}
The group performed an online survey of the U.S. particle physics community to gain background on how practitioners in the field at all levels assessed their own physics education and training, with the aim to identify areas of challenge and  inform potential improvements for the future.  The group's report ~\cite{cp3} is detailed and replete with findings and recommendations, from which we have identified the following succinct elements:
\begin{enumerate}
\setcounter{enumi}{3}
    
    \item Graduate programs in particle physics should normalize training for a broad range of STEM careers with appropriate formal courses, rather than forcing students to resort prematurely to self-teaching or peer learning. The courses should provide strong grounding in particle physics and mathematics, but also computation, statistics and instrumentation, with the aim to benefit careers in physics, industry or education. 
 
    \item Correspondingly, Universities should provide undergraduate students with a more complete picture of what particle physics trained researchers do. A realistic view of common career paths post baccalaureate and post graduate school should be presented, including for theoretical and experimental positions as well as non-academic careers.

    \item Undergraduate participation in the survey was rather minimal, driven in part by lack of information.  With support from Professional Societies and Physics Departments, connections and networking opportunities should be developed to strengthen connections for undergraduate students with HEPA community activities.  And the HEPA Community should actively plan to perform a future survey with emphasis on undergraduate participation, to assess and strengthen links to students at that academic level.
    .
 \end{enumerate} 
 
\subsection{A New Perspective on the Masters Degree}
As noted in several of the recommendations derived from the comprehensive survey cited above and detailed in the Contributed Paper ~\cite{cp3},  a  potentially important professional level post baccalaureate, the Masters Degree, is often overlooked or ignored by the physics community which tends to be "Ph D" driven.  The Masters level has several important attributes worthy of community consideration.  (1) It is a level where more extensive cross-disciplinary (elective) course work is possible, providing potential branches to applied math, statistics, computer science, engineering and nuclear medicine. (2) It is a level of potential participation by and engagement with students from the private sector who enroll for technical advancement with support from their companies.  This nexus can provide a bridge between students following an academic path with those already in commercial applications sectors, opening up dialog and potential career opportunities that might otherwise be overlooked.  (3) The Masters Level also affords an intermediate (and potentially achievable) academic target for students from groups traditionally underrepresented in physics and for whom a PhD in physics might seem an unlikely goal.  And (4) enriched programs at the Masters Level can lead to collaborative opportunities across academia, which is the central topic of the next section.
\begin{enumerate}
\setcounter{enumi}{6}

    \item Universities, especially also non-research universities, should consider setting up Masters Degree programs in particle physics and related areas, such as hardware and software technology for Big Science experiments.
\end{enumerate}
    
\section{Collaborative Opportunities Across Academia}   
\label{section:Section4}

The emphases of this group were to review the challenges and develop strategies to help transform the particle physics field into a stronger and more diverse ecosystem of talent and expertise, with the expectation of long-lasting scientific and societal benefits. Regarded as central were the  building of  collaborative bridges between faculty and experimental programs at R1 institutions (major research universities and national laboratories) with those at R2 institutions (that provide training up to masters degrees), Predominantly Undergraduate Institutions (PUIs) and Community Colleges (CC)). The intended aims over the next decade are to:

\begin{enumerate}
\setcounter{enumi}{7}
\item Expand the benefits of faculty collaboration and research opportunities across the broad spectrum of academia and give equivalence: opportunities for all in technical and scientific leadership on projects and with appropriate recognition for contributions. 

\item In corollary, conduct a study of new models of collaboration or cooperation that would allow R2/PUI/CC faculty and their students to collaborate in demonstrably effective ways in experiments.  This includes addressing the challenges of teaching loads, student training and funding availability that directly impact consequential participation.

\item Broadly accessible data and analysis platforms will benefit student access and  participation.  In the true spirit of Open Science, the HEP community should define, with cogent arguments, what should be the scope of making our data and resources publicly available, and the hardware, software and person-power costs associated with such implementation. 

\end{enumerate}

Machine Learning (ML) is becoming an integral part of physics research. Many critical HEP algorithms for triggering, reconstruction, and analysis rely on ML and there are entire conferences and summer schools dedicated to this crosscutting field. Despite the relevance and importance of this research, pursuing a career at the intersection of these fields remains a tenuous and undefined endeavor. The current mindset in the field is that highly specialized skills such as software and firmware development and instrumentation development are not broadly recognized as “physics” work.  This perception of what it means to ‘be a physicist’ must be challenged lest it continue to be an impediment. Confronting and addressing these issues would encourage an influx of new workforce into the field, help retain those who are in the field, and equip those who might seek careers outside of HEP.
\begin{enumerate}
\setcounter{enumi}{10}

\item Qualification for HEP faculty jobs should not be based solely on physics analysis but must be expanded to to include computing, software and/or hardware contributions.  

\end{enumerate}

\section{International Opportunities for Particle Physics Education}   
\label{section:Section6}
Particle physics is a global endeavor. No one institution or nation can assemble the resources or expertise needed to explore the frontiers of the field. The diversity of national, social and cultural backgrounds present in the experiments and labs enriches the pool of intellectual thought and solidifies the validity of their scientific findings.  And no one country or region has a monopoly on good ideas or approaches to strengthen the educational experience for studwnts.  But there are noteworthy examples of excellent ongoing efforts, which this working group reviewed in detail. Their line of thought has been worked out in a contributed paper~\cite{CP5}
and hasled to several noteworthy recommendations.
\begin{enumerate}
\setcounter{enumi}{11}
    \item U.S. based pre-university particle physics collaborations, such as QuarkNet and other outreach programmes, should expand collaboration with international partners, such as the International Particle Physics Outreach Group (IPPOG), the CERN Beamline for Schools (BL4S) and Teacher summer school programmes in Europe, and should collaborate with partners in the developing world, such as the African School of Fundamental Physics and Applications. The participation in the Global Cosmics portal should be enhanced by developing low-cost cosmic ray detectors for educational use.
    \item Student exchange programs should be fostered and supported, such as the NSF Research Experience for Undergraduates (REU) program, which funds participation of U.S. students in the CERN Summer Student program, and the DoE-INFN summer student exchange program between the U.S. and Italy. Where possible these should be extended, in particular with student exchange programs and summer schools in developing countries, such as the African School of Fundamental Physics and Applications.
\end{enumerate}
\section{Interconnections and Synergies with Community Engagement Topical Groups}   
\label{section:Synergies}
Physics Education (CEF4) has important interconnections to other Snowmass 2021 Topical Groups within the Community Engagement Frontier. 

\textbf{With CEF1, the Topical Group on Applications and Industry:}. Opportunties are potentially vast, linking education with technological innovation.  National Laboratories are a natural training ground for student research opportunities. And at the Masters Degree level, universities and colleges are a fertile training ground to mix students moving up to STEM careers with students from industry who seek to further enrich their technological training.  This is a melting pot worth pursiuing energetically and behooves universities and colleges to consider the Masters Degree in Physics in a new light.

\textbf{With CEF2, the Topical Group on Career Pipeline and Development:} Physics Education (CEF4) is central to creating a skilled workforce pipeline to all HEP Frontiers and beyond for STEM areas in industry. All recommendations in CEF2 are strongly endorsed by this CEF4 Working Group. Beyond regular course curriculum, Software Training programs and Open Science activities~\cite{cp14} can go a long way to attracting talent in HEP as well as preparing HEP talent for STEM industry jobs. Strong participation by faculties and students of PUIs and CCs in HEP programs~\cite{cpex2} will enable a diverse and inclusive STEM workforce for HEP and industry. 

\textbf{With CEF3, the Topical Group on Diversity and Inclusion:}  With the inclusion of a well-educated and demographically and geographically diverse group of students, the person power needed for current and future experimental and scientific challenges will be in place to meet the needs of particle physics community over the next two decades.

\textbf{With CEF5, the Topical Group on Public Education and Outreach:} K/12 education is fundamentally public education at its most fundamental level.  With students excited and engaged, and their bringing that interest and excitement home to their families, the thrill of particle physics and the possibilities of future discoveries will help build the strong supportive community needed to sustain the science.

\textbf{With CEF6, the Topical Group on Public Policy and Government Engagement:} Particle Physics is an exciting, yet esoteric field.  It is therefore  challenging to translate that excitement to those who make public policy and provide funding support, when they are under myriad pressures from across the political spectrum.   However, hand-in-hand with the expansive educational opportunities that discovery science affords, and through mention of the excitement and participation of students across the nation, the conversational optics can be refocused in constructive ways, to the benefit of the field and nation.

\textbf{With CEF7 the Topical Group on Environmental and Societal Impacts:} At the time of Fermilab's founding near the village of Weston, Illinois, the land was prairie.  Since that time, the laboratory has had programs on environment and ecology that are particularly accessible to younger students and to engage them in the natural world.  These programs, coordinated through the Lederman Science Center at Fermilab, are a wonderful model of the interconnection of particle physicists, education, and local schools and  communities~\cite{prairie},~\cite{prairiekids}.  While there are outreach aspects to such programs, the latter reference is intrinsically educational by design.
\section{Conclusions}   
\label{section:Conclusions}

To support a compelling program of scientific discovery in the long run, a robust education program in physics, mathematics and the sciences is an essential companion. Such a program should provide students across the demographic spectrum ample basis of opportunity to enter particle physics and ancillary fields to engage in and benefit from the science.

The ten-year program and twenty-year vision of Snowmass 2021 affords a strategic opportunity and time window within which both the science and the education process can evolve holistically and constructively. \textbf{This requires Community Engagement at all educational levels, which can only be sustained when the people involved are appropriately recognised, credited and supported for their contributions.}

\clearpage

\bibliographystyle{JHEP}

\bibliography{bibliography}  
\end{document}